# Monitoring of Optical Networks Using Correlation-Aided Time-Domain Reflectometry with Direct and Coherent Detection

Michael H. Eiselt
Adtran Networks SE
Meiningen, Germany
meiselt@adva.com

Florian Azendorf
Adtran Networks SE
Meiningen, Germany
fazendorf@adva.com

André Sandmann
Adtran Networks SE
Meiningen, Germany
asandmann@adva.com

Florian Spinty
Adtran Networks SE
Meiningen, Germany
fspinty@adva.com

Mirko Lawin
Adtran Networks SE
Meiningen, Germany
mlawin@adva.com

*Abstract*—We report on methods to monitor the transmission path in optical networks using a correlation-based OTDR technique with direct and coherent detection. A high probing symbol rate can provide picosecond-accuracy of the fiber propagation delay, while a sensitive phase detection with a high repetition rate allows the monitoring of dynamic effects in the vicinity of the fiber. We discuss various approaches to evaluate the measured traces and show the results of a few monitoring applications.

## I. Introduction

Increasing data volume transmitted over optical networks increases the impact of failures in the physical network layer, i.e. the transmission fiber. Furthermore, recent attacks on communication networks have shown that also other critical networks, like the railroad or the power network, depend on a functioning telecommunication network. To protect the optical network against accidental or intentional failures, monitoring the integrity of the transmission fiber using optical time domain reflectometry (OTDR) has long been used and has recently become widespread by integration into network equipment.

However, OTDR only monitors the attenuation of the transmission fiber and can detect problems only after they occur. Imminent damage, like fiber strain due to broken tree branches lying on a pole mounted fiber, can rarely be detected. Also other effects, like changes to the fiber propagation delay, potentially impacting the network synchronization, can only coarsely be detected. Therefore, more sophisticated fiber monitoring techniques have been recently investigated, which were previously known only for sensing applications, like the distributed acoustic sensing (DAS). Using this technique, a wider variety of effects can be detected [1].

Going a step further, the information gathered about the fiber during monitoring can in some cases be used as an indicator for environmental parameters. These parameters might be the outside temperature or strain to the fiber, but also acoustic vibrations. This additional information can be used

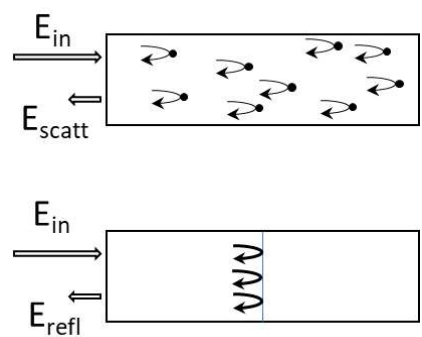

Fig. 1. Top: Schematic of backscattering of input field $E_{in}$ at impurities in the fiber. Scattered field components with random phases add coherently to $E_{scatt}$. Bottom: Schematic of reflection of input field $E_{in}$ at a splice in the fiber. Reflected field components add with the same phase

for other purposes like traffic monitoring [2] or submarine earthquake detection [3].

In this paper, we will discuss modifications of a standard OTDR by using correlation to improve the spatial resolution and coherent detection to enhance the receiver sensitivity and to gather additional information on the phase of the backscattered signal. As a widespread deployment of network monitoring equipment requires a limited cost, it is an objective to use telecommunications grade equipment, which is manufactured in high volume, contributing to a reduction in cost over dedicated sensing-application equipment.

## II. Correlation to Improve Spatial Resolution

For optical time-domain reflectometry, an optical pulse is transmitted into the monitored fiber. During the propagation in the fiber, two effects lead to the return of a part of the forward travelling pulse power back to the fiber input. First, in each fiber section, small impurities in the fiber scatter a small fraction of the optical power, of which a part is captured by the fiber in backward direction (see Fig. 1, top). This Rayleigh backscattering is random due to the random localization of the impurities and the coherent superposition of the backscattered fields. In a standard transmission fiber, the typical backscatter coefficient is approximately -70 dB/m. Second, at transitions between different refractive indices in

The work was carried out in part in the SoFiN project, funded by the European Union under grant 101039105. Views and opinions expressed are however those of the authors only and do not necessarily reflect those of the European Union or the European Health and Digital Executive Agency. Neither the European Union nor the granting authority can be held responsible for them.

the propagation path, the forward travelling power experiences Fresnel reflections (see Fig. 1, bottom). These transitions could be small air gaps in connectors, but also slightly different fiber types. Depending on the connector type and the quality grade, the minimum return loss is 26 to 45 dB for a physical contact (PC) connector and 55 dB for an angled physical contact (APC) connector [4], while the return loss of a fusion splice should be larger than 60 dB [5].

The backscattered and reflected power is received, and the time evolution of the received power gives information on the spatial power evolution along the fiber, including higher-loss sections and fiber breaks. In general, the spatial resolution of an OTDR is limited by the width of the optical pulse transmitted into the fiber. The longer the length of the fiber under test, the longer the pulse must be (at constant pulse amplitude) to achieve a sufficient signal-to-noise-ratio (SNR) at the receiver, and the worse becomes the spatial resolution. If, however, a modulated bit sequence is transmitted instead of a single long pulse, the backscattered signal can be cross-correlated with the transmitted sequence, such that the overall accumulated power is approximately the same as if a pulse with the length of the sequence were transmitted, but the spatial resolution corresponds to the length of a single bit of the sequence [6]. By using a high bit rate for the probe signal, a spatial resolution of sub-mm (less than 10 ps round-trip propagation time) can be achieved even over a fiber length of 100 km. This performance, however, not only requires low-noise optical signals, but also a clock with an accuracy of a few ppb to measure the propagation time [6].

### III. Direct detection correlation OTDR

The ability to measure the propagation delay of a transmission fiber with a picosecond accuracy allows a precise synchronization between two clocks in a telecommunication system. For example, the often used protocol IEEE 1588v2, or precision time protocol (PTP) [8], relies on a symmetric delay between the master clock and the slave clock in both transmission directions. Any asymmetry in the propagation delay leads to an error in the slave clock synchronization. Symmetry requirements are on the order of a few nanoseconds. Determining and tracking any asymmetry increases the quality of the synchronization. Propagation delay in deployed fiber can change due to variations in the fiber temperature. Typically, the change of refractive index is the dominant effect, leading to a propagation time dependence of approximately 35 ps/(K*km) [9]. Therefore, the propagation delay change can also be used to estimate the environmental temperature of the fiber. Using the correlation OTDR technique, in [10], the evolution of the propagation delay in 8.5 km deployed fiber was measured over 14 days and compared to the air temperature changes. It turned out that the measured delay changes fit very well to the temperature changes, if a heating / cooling time constant of the 1-meter deep buried fiber of 12.7 days was taken into account, with which the fiber temperature followed the air temperature. Over a period of one year, delay variations of more than 10 ns were measured [6].

These measurements were performed with a $2^7-1$ PRBS with a data rate of 10 Gbps as the probe sequence. At the receiver, the signal was sampled at 50 GSps to yield, after pulse fitting, an accuracy of approximately 2 ps. The setup was subsequently implemented using an SFP+ module as optical transceiver and a Xilinx multi-processor system-on-chip (MPSoC) for signal sampling and data processing [11].

Analog-to-digital conversion was performed by a 1-bit slicer with averaging of 4000 traces to increase the receiver resolution. This demonstrates how the C-OTDR system can be implemented by using low-cost telecommunication components.

### IV. Coherent detection for sensitivity and resolution improvement

A drawback of direct detection is the fact that closely spaced reflections lead to a coherent superposition of the reflected optical fields, which are detected in the receiver via the non-linear function of the photodiode, such that a cross correlation with the transmitted sequence does not lead to two discrete correlation peaks. The round-trip time between two reflections should therefore be longer than the sequence duration (e.g. 100 ns / 10 m for a 1000-bit sequence at 10 Gbps). This requirement can be alleviated by using coherent detection in the receiver, where the superimposed reflection signals are added linearly and can be distinguished by cross-correlation. A further advantage of coherent detection is the amplification which the detected signal experiences from the local oscillator. Furthermore, phase modulation can be used for the probe sequence, e.g. binary phase-shift keying (BPSK), further improving the receiver sensitivity.

The improved minimum spacing between reflections has been demonstrated in [12], where 2000 Bragg gratings, written into a fiber with a spacing of 50 mm, were interrogated using a 2048-bit sequence at 5 Gbps as probe sequence for a coherent correlation OTDR. The bit length of 200 ps corresponds to the round-trip time over 20 mm and thus determines the resolution of the measurement. As the spacing of the gratings corresponds to a round-trip time of 500 ps, more than 800 reflections of the 410-ns long sequence superimposed coherently at the receiver, but the individual reflections could still be clearly separated. In subsequent measurements, the probe wavelength was tuned in steps of 8 pm, yielding the reflection spectrum of each Bragg grating. This measurement revealed that the strain from spooling the fiber resulted in a detuning of the Bragg wavelengths by 150 pm, periodic along the fiber with a period of the spool circumference of approx. 50 cm.

### V. Coherent detection with phase evaluation

The phase acquired by the probe signal when propagating along the fiber is strongly impacted by external effects, which modify the refractive index of the fiber. For example, the temperature dependence of the refractive index mentioned above leads to a refractive index change of approximately $10^{-6}$ for a temperature change of 0.1 K. Over a length of 10 cm, this results in a phase change of 0.4 rad for a probe wavelength of 1550 nm. The phase change can be probed in two different ways. First, the backscattering components of a probe pulse are coherently combined with different phases such that the amplitude of the backscattering is varied. This can be called the local coherence effect. Secondly, when the reflected or backscattered probe pulse is detected in a coherent receiver and superimposed with a local oscillator, the measured phase reveals the phase which the probe pulse acquired during propagation along the whole sensing fiber. Any change of phase at some location in the fiber can be seen in the receiver. This can be called the remote coherence effect. The remote coherence effect is typically stronger than the local effect, as external effects that impact a longer section of the fiber (e.g. strain from acoustic waves) are probed along the whole impacting length, whereas the local effect is accumulated only along the round-trip length of the probe pulse duration.

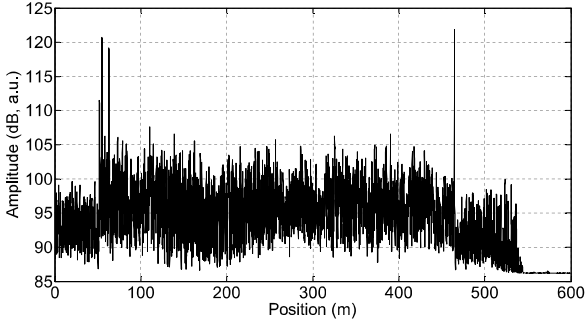

Fig. 2. Backscttering trace ("fingerprint") of a 400-m long fiber, showing reflections from connectors at the fiber input and output.

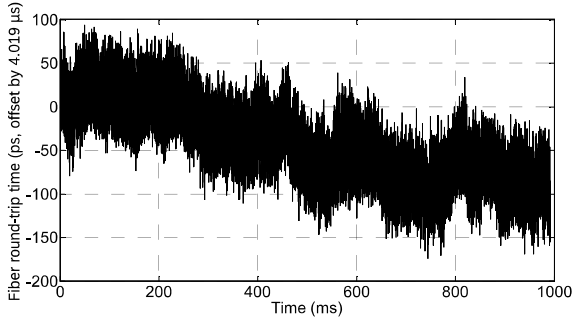

Fig. 3. Evolution of fiber round-trip time between input and output connectors over a time period of one second.

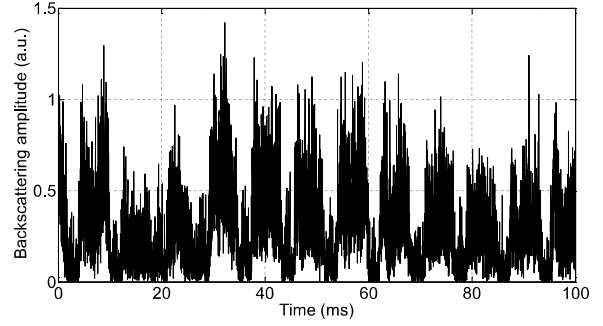

Fig. 4. Time variation of backscattered power from a point 198.3 m into the fiber, showing a 120-Hz component from acoustic tone.

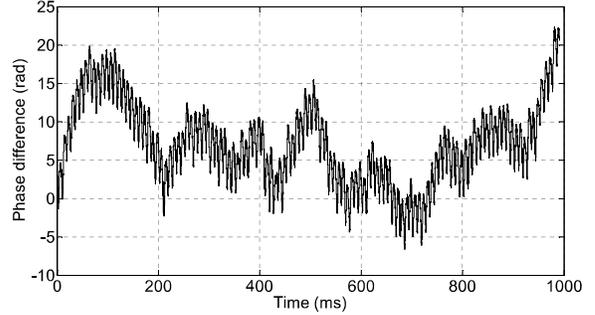

Fig. 5. Variation of round-trip propagation phase between input and output connectors of the 400-m long fiber.

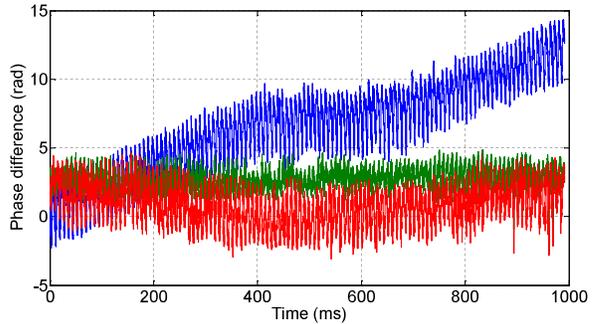

Fig. 6. Variation of round-trip propagation phase in three neighboring sections of between 2 m and 3.5 m length, starting at 193.9 m (blue), 195.9 m (green) and 199.4 m (red) from the fiber input connector.

While the local effect is often used in distributed acoustic sensing (DAS) applications, receiving the phase of the backscattered signal has, for instance, been demonstrated in an experiment, where the differential phase variations along a fiber were evaluated to detect a speech signal [13].

## VI. EVALUATION OF BACKSCATTERING TRACES

As has already been mentioned above, the recorded backscattering traces can be evaluated in various ways. The example considered below is based on an approximately 400 m long fiber, which is probed using a 256-bit extended PRBS sequence at a data rate of 320 Mbps, sampled at the receiver at 1.6 GSps. Around 200 m into the fiber, 2 meters of the fiber are stretched and exposed to a 120-Hz acoustic signal.

The power of the backscattered signal after de-correlation, averaged over a second, i.e. the "fingerprint" of the fiber, is shown in Figure 2. The larger peaks stem from reflections at FC/PC connectors. The signal after the fiber end peak stems from side lobes of the cross-correlation over a range of 256 bits of the probe signal, followed by the thermal noise level.

To monitor the impact of external effects on the sensing fiber the individual finger prints can be evaluated. In a first step, the round-trip propagation time between the input and output reflections can be monitored with an accuracy better than 10 ps for a data rate of 10 Gbps with a sampling rate of 50 GSps [6]. Figure 3 shows the time evolution of the roundtrip time (RTT) to the end reflector. While the RTT is around 4.0 µs for the 400 m long fiber, the variations over one second are on the order of about 100 ps. The variations do, however, not resemble the 120-Hz acoustic signal imposed on a section of the fiber. As the sampling rate was only 1.6 GSps, the accuracy after pulse fitting was reduced to approximately 100 ps, as seen by the variations of the curve in Figure 3.

In a second step, the local effect of the acoustic signal can be observed, as it changes the backscattering from a particular location in the fiber. In Figure 4, the variation of the backscattered power over time is shown for a point at 198.3 m from the input fiber connector. It can be seen that the backscattering varies with a frequency of 120 Hz. However, this behavior can be observed only for a few backscattering points in the fiber section, while others do not exhibit this frequency.

While so far only the amplitude of the backscattered or reflected signals have been used for the evaluation of external impacts, in a third step the phase evolution of the optical probe signal along the fiber is considered. While the variation of the phase of a received signal depends strongly on the phase noise of the probe laser, the phase difference between the backscattered signals from two points in the fiber cancels most of the phase noise effect. Therefore, we can monitor the evolution of the propagation phase in a section of the fiber by measuring the evolution of the phase difference between backscattering locations before and after the fiber section.

Figure 5 shows the phase evolution over the whole 400-m length of the fiber by taking the phase difference between the signals reflected at the connectors at the fiber input and the fiber output. Here the 120-Hz acoustic signal can clearly be seen, introducing an approximately 7-radians peak-peak modulation of the probe phase, in addition to a slower phase modulation, potentially resulting from thermal and other acoustic effects. A position resolved variation can be obtained by considering the evolution of the phase difference between closely spaced backscattering points. Here, the phase extraction is noisier, as the backscattering amplitudes are much lower than the reflection amplitudes. Figure 6 shows the evolution of the round-trip propagation phases over three fiber sections between backscattering points at 193.9 m, 195.9 m, 199.4 m, and 202.4 m from the fiber input connector. Besides slowly varying phase differences, peak-to-peak variations between 1.5 and 4 radians can be observed for each curve.

VII. SUMMARY

Techniques using optical time-domain reflectometry can be used to measure variations in the refractive index of an optical fiber, which can be attributed to a variety of external effects. Depending on the required position resolution and sensitivity, either the phase or the amplitude of the backscattered or reflected signal can be evaluated. This gives a toolbox to monitor the fiber for a variety of effects, reaching from slow effects like temperature or strain to effects in the kilohertz range like seismic or acoustic vibrations. Telecommunications grade components can be used for the implementation of the interrogators, enabling a high-volume, low-cost deployment of optical network monitors.